\documentclass[aip,apl,reprint]{revtex4-1}

\usepackage{graphicx}
\usepackage{dcolumn}
\usepackage{bm}
\usepackage{amsmath}

\begin{document}

\title{Immobilization of a bubble in water by nanoelectrolysis}

\affiliation{CINaM-CNRS Aix-Marseille Univ., Campus de Luminy, case 913, 13288 Marseille Cedex 9, France}

\author{Zoubida Hammadi}

\author{Laurent Lapena}

\author{Roger Morin}

\author{Juan Olives}
\email{olives@cinam.univ-mrs.fr}

\date{\today}

\begin{abstract}
A surprising phenomenon is presented: a bubble, produced from water electrolysis, is immobilized in the liquid (as if the Archimedes' buoyant force were annihilated). This is achieved using a nanoelectrode (1 nm to 1 $\mu$m of curvature radius at the apex) and an alternating electric potential with adapted values of amplitude and frequency. A simple model based on ``nanoelectrolysis'' (i.e., nanolocalization of the production of $\rm H_2$ and $\rm O_2$ molecules at the apex of the nanoelectrode) and an ``open bubble'' (i.e., exchanging $\rm H_2$ and $\rm O_2$ molecules with the solution) explains most of the observations.
\end{abstract}

\maketitle

Microbubbles have many applications: medicine (contrast agents,\cite{Bloch-etal:2004} gas embolotherapy,\cite{Qamar-etal:2010} blood clot lysis \cite{Acconcia-etal:2013}) or nuclear industry.\cite{Kim-etal:2000} Questions about their stability remain topical.\cite{Aoki-etal:2012} $\rm H_2$/$\rm O_2$ microbubbles are easily generated by water electrolysis and their production is controlled by the electric potential. Recently, $\rm H_2$ microbubbles have been used to rotate a microobject.\cite{Li-and-Hu:2013} Droplets may also be controlled by electrolysis.\cite{Poulain-etal:2015} Many applications will result from a more effective control on the microbubbles and, especially, on each individual microbubble. Our preceding paper\cite{Hammadi-etal:2013} showed a first example of such control: the nanolocalization of the production of microbubbles at a unique point of a tip-shaped electrode, under precise values of the amplitude and the frequency of the alternating potential. This phenomenon will be called nanoelectrolysis (for bubble production). The relation between microbubble production and current intensity will be given in a forthcoming paper. The present paper describes a surprising phenomenon, which represents an example of control of a single microbubble: the immobilization of a bubble in the liquid (as if the Archimedes' buoyant force were annihilated).

\begin{figure}[t]
\begin{center}
\includegraphics[width=2.4cm]{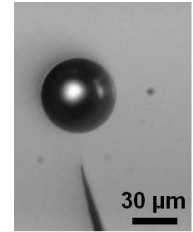}
\end{center}
\caption{Microbubble immobilized in the solution, above the apex of the nanoelectrode (potential $V_{\rm m}$ = 3 V, $\nu$ = 1000 Hz).} \label{Bubble}
\end{figure}

\begin{figure}[t]
\begin{center}
\includegraphics[width=7.5cm]{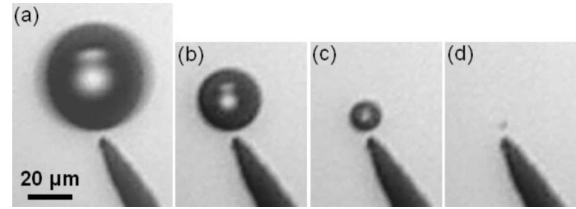}
\end{center}
\caption{Immobilized bubble, slowly decreasing with time ($V_{\rm m}$ = 3.43 V, $\nu$ = 1000 Hz): $t$ = 0 (a), $t$ = 2 min 6~s (b), $t$ = 3 min 55 s (c) and $t$ = 4 min 33 s (d). Same scale for all the images.} \label{Bubble-decrease}
\end{figure}

\begin{figure}[t]
\begin{center}
\includegraphics[width=5.5cm]{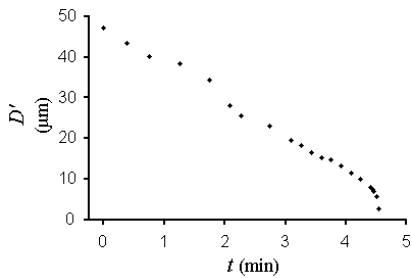}
\end{center}
\caption{Diameter $D'$ of the bubble versus time $t$, for the experiment shown in Fig.~\ref{Bubble-decrease}.} \label{Diameter}
\end{figure}

\begin{figure}[t]
\begin{center}
\includegraphics[width=5.1cm]{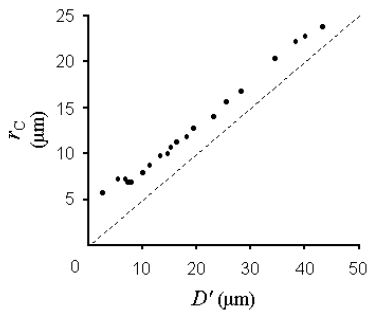}
\end{center}
\caption{Relation between $r_{\rm C}$ (distance between the center of the bubble and the apex of the electrode) and $D'$ (diameter of the bubble), for the experiment shown in Fig.~\ref{Bubble-decrease}. The dashed line represents the equation $r_{\rm C} = D'/2$.} \label{Center}
\end{figure}

\begin{figure}[t]
\begin{center}
\includegraphics[width=7cm]{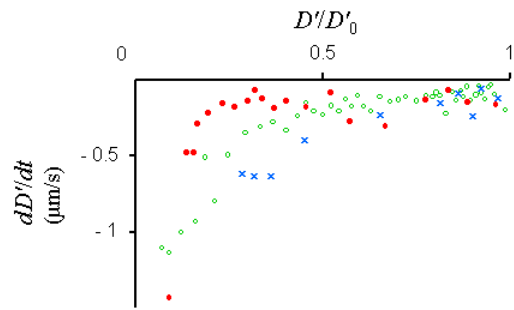}
\end{center}
\caption{Diameter decreasing rate $dD'/dt$ versus $D'/D'_0$ ($D'_0$ = initial diameter at $t$ = 0) for the experiment shown in Fig.~\ref{Bubble-decrease} (solid circles), another experiment with $V_{\rm m}$ = 27.5 V, $\nu$ = 500 Hz, and $D'_0$ = 63 $\mu$m (crosses), and a third experiment with $V_{\rm m}$ = 21 V, $\nu$ = 500 Hz, and $D'_0$ = 260 $\mu$m (empty circles).} \label{Decreasing-rate}
\end{figure}

The experimental procedure is described in Ref.~\onlinecite{Hammadi-etal:2013}. Water electrolysis is performed using an aqueous solution containing $10^{-4}$ mol/L of $\rm H_2SO_4$ and a periodic alternating applied potential, here of rectangular shape:\footnote{A potential of sinusoidal shape may also be used.} $V(t) = V_{\rm m}$ for $0<t<T/2$, $V(t) = -V_{\rm m}$ for $T/2<t<T$ ($T$ is the period and $\nu=1/T$ the frequency). The amplitudes $V_{\rm m}$ typically range from 2 V to 30 V. Two Pt electrodes are used and one of the two electrodes---called nanoelectrode---is tip-shaped, with a curvature radius, at the apex of the electrode, ranging from 1 nm to 1 $\mu$m. In the previous paper,\cite{Hammadi-etal:2013} we showed that, for definite values of the amplitude and the frequency of the potential, the microbubbles are produced at a single point, the apex of the nanoelectrode. We start in such conditions, so that microbubbles are successively generated at the apex of the electrode and then naturally go up towards the air--solution surface, owing to the Archimedes' buoyant force. Let us now describe the observed phenomenon. Immediately after the production of a microbubble, if we rapidly increase the frequency (generally, up to at least 500 or 1000 Hz; and then, maintaining constant the frequency; $V_{\rm m}$ remaining always constant), then the microbubble remains immobile in the solution,\footnote{If the increase in frequency is not sufficiently rapid, the bubble will continue to go up and will not be immobilized.} at some distance from the electrode and nearly at the vertical of the apex of the electrode (Fig.~\ref{Bubble}).
In addition, this configuration is stable, since the bubble remains at the vertical of the apex of the electrode and at the same distance from the electrode, when this electrode is moved in any direction with respect to the solution (see the video\cite{Supplement}). This immobilization is observed during various minutes or hours (at constant amplitude $V_{\rm m}$ and frequency). The diameters $D'$ of the immobilized bubbles typically range from some $\mu$m to 400 $\mu$m. This diameter may remain constant, in some cases, but very frequently decreases with time, as shown in Fig.~\ref{Bubble-decrease}, till the (optical) disappearance of the microbubble.
As the diameter $D'$ decreases with time (Fig.~\ref{Diameter}), we observe that the distance $r_{\rm C}$ between the center of the bubble and the apex of the electrode also decreases (Fig.~\ref{Center}).
The evolution of the diameter decreasing rate $dD'/dt$ as a function of the diameter is shown in Fig.~\ref{Decreasing-rate}, for three different experiments.

In order to understand such a strange phenomenon, we first note that the bubble cannot be considered as a closed system (i.e., with no exchange of matter with the solution) and its immobilization as a balance between the Archimedes' buoyant force and a new force. Indeed, such a (downward) force should increase with the distance between the bubble and the apex of the electrode (in order to explain the stability of this equilibrium, shown in the video\cite{Supplement}), which seems unrealistic. Our explanation is then based on a bubble considered as an open system, exchanging matter with the solution through its (immobile) surface. In this case, the new forces which might counterbalance the Archimedes' one are the flux of momentum entering the bubble and the hydrodynamic forces. However, these forces are intimately associated with the fields of the pressures, velocities, and densities and the determination of these fields is a very complex coupled problem (involving hydrodynamics, diffusion, and mass fluxes kinetics at the bubble--solution interface). This problem is not treated here and will be investigated in the future by using computer simulations. As an approximation, hydrodynamics equations (concerning the pressures and the velocities) may be separated and, in the following, we show that a simplified model based on the mass balance, diffusion, and interface fluxes equations may explain---at least qualitatively---most of the above observations.  

In our previous paper,\cite{Hammadi-etal:2013} we showed that the bubble production is nanolocalized (at the apex of the nanoelectrode) when the potential $v$ (applied to the dielectric layer, at the electrode--electrolyte interface) reaches a threshold value $v_0$. Similarly, we expect that the chemical reactions of electrolysis will be also nanolocalized for a (lower) threshold value $v'_0$. According to our model,\cite{Hammadi-etal:2013} this implies that, in the potential amplitude--frequency plane, the domain for the nanolocalization of the electrolysis reactions is shifted toward higher frequencies, with respect to that for the nanolocalization of bubble production. Thus, after the increase in frequency needed for the immobilization of the bubble, the new present amplitude--frequency values correspond to conditions of no bubble production,\cite{Hammadi-etal:2013} but we here assume that they correspond to nanolocalization conditions for the electrolysis reactions: i.e., these reactions still occur and are nanolocalized at the apex of the electrode, although the rate of production of $\rm H_2$ and $\rm O_2$ molecules is not enough to generate bubbles. These molecules, produced at the apex of the electrode, thus remain and diffuse in the solution, without generating any bubble. In our model, the immobilized bubble is considered as an open system, which may exchange $\rm H_2$ and $\rm O_2$ molecules with the solution through its surface. The whole system (solution and bubble) is considered in a steady state.

The bubble contains $\rm H_2$ and $\rm O_2$ molecules, and the general model for a bubble containing these two components is given in the supplementary material.\cite{Supplement} This model shows that, in the steady state (and for not too small diameters), the bubble contains nearly the same number of moles of $\rm H_2$ and $\rm O_2$ (although the production of $\rm O_2$ molecules is lower than that of $\rm H_2$, and the solubility of $\rm O_2$ is higher than that of $\rm H_2$; this is due to the low diffusion coefficient of $\rm O_2$).\cite{Supplement} In the following, for the sake of simplicity, we present the model for a bubble containing only one component, e.g., $\rm H_2$.  Let us use the subscript $i$ = 1 to denote the $\rm H_2$ component in the solution and the subscripts $i$ = 2, 3, ... for the other components of the solution (such as $\rm O_2$, $\rm H_2O$, $\rm H^+$, ...). The diffusion flux of $\rm H_2$ is given by Fick's law:
\begin{eqnarray}
j_1 = \rho_1({\rm v_1} - {\rm v}) = -D_1\,{\rm grad}\rho_1 \label{Fick}
\end{eqnarray}
($\rho_i$ and ${\rm v}_i$ are, respectively, the volume mass density and the velocity of the component $i$; $\rm v$ is the barycentric or convection velocity defined by $\rho\,{\rm v} = \rho_1\,{\rm v_1} + \rho_2\,{\rm v_2} +...$, $\rho$ being the mass density of the solution; $D_1$ is the diffusion coefficient $= 4.6 \times 10^{-9}\; \rm m^2/s$ for $\rm H_2$ in water at $20^{\circ}$C). In the solution, the $\rm H_2$ mass balance equation
\begin{eqnarray}
\frac{\partial \rho_1}{\partial t} + {\rm div}(\rho_1\,{\rm v_1}) = \dot{m_1} \delta \label{Mass}
\end{eqnarray}
($\dot{m_1}$ is the mass production per unit time at the apex of the electrode, which is considered as the origin point; $\delta$ is the Dirac measure---in space---at this origin point; $\dot{m_1}$, averaged on a period, is considered constant in this simple model) leads to
\begin{eqnarray}
- D_1 \Delta \rho_1 + {\rm v}\cdot{\rm grad}\rho_1 = \dot{m_1} \delta \label{Diffusion-convection}
\end{eqnarray}
(with the help of Eq.~(\ref{Fick}), $\frac{\partial \rho_1}{\partial t} = 0$---steady state---and ${\rm div}\,{\rm v} = 0$---incompressibility). 

At each point of the surface $\rm S$ of the bubble, the flux of $\rm H_2$ entering the bubble is
\begin{eqnarray}
j_{\rm S} = \rho_1({\rm v_1} - {\rm v_S})\cdot{\rm n} 
= \rho({\rm v} - {\rm v_S})\cdot{\rm n} \label{Flux}
\end{eqnarray}
(assuming only exchanges of $\rm H_2$ between the bubble and the solution; $\rm n$ is the unit vector normal to $\rm S$, oriented towards the interior of the bubble; $\rm v_S$ is the normal velocity of the surface, parallel to $\rm n$; obviously, $\rm v_S = 0$ for an immobile bubble), which gives, according to Eq.~(\ref{Fick})
\begin{eqnarray}
-D_1\,\partial_{\rm n}\rho_1 = (1 - \frac{\rho_1}{\rho})j_{\rm S} \approx j_{\rm S}. \label{S-condition}
\end{eqnarray}
We assume that the exchange of $\rm H_2$ between the solution and the bubble is controlled by the simple kinetic law
\begin{eqnarray}
j_{\rm S} = K(\rho_1 - H\, p'), \label{Flux-law}
\end{eqnarray}
where $p' = p_{\rm a} + \frac{2\gamma}{R'}$ is the pressure in the bubble ($p_{\rm a}$ the atmospheric pressure, $\gamma$ the surface tension, $R' = D'/2$ the bubble radius), $H$ the Henry's constant, and $K$ the mass transfer coefficient ($\gamma = 7.28 \times 10^{-2}$ N/m, the water--air value at $20^{\circ}$C; $H = 1.62 \times 10^{-3}$ kg/($\rm m^3$~atm) for $\rm H_2$ in water at $20^{\circ}$C (Ref.~\onlinecite{Sander:2015})).

As a simple model, we use the $\rho_1$ field produced only by diffusion (i.e., satisfying to Eq.~(\ref{Diffusion-convection}) with ${\rm v} = 0$), without taking into account the presence of the bubble (i.e., the condition Eq.~(\ref{S-condition}))
\begin{eqnarray}
\rho_1 = \frac{\dot{m_1}}{4\pi D_1}\,\frac{1}{r}, \label{rho1}
\end{eqnarray}
where $r$ is the distance to the origin point $\rm O$ (which is the apex of the electrode). If $r_0$ denotes the distance at which $\rho_1(r_0) = H\,p_{\rm a}$ (corresponding to the equilibrium of $\rm H_2$ between the solution and $\rm H_2$ gas at the atmospheric pressure), Eq.~(\ref{Flux-law}) takes the form
\begin{eqnarray}
j_{\rm S} = KHp_{\rm a}(\frac{r_0}{r} - (1 + \frac{2\tilde{\gamma}}{R'})), \label{Flux-expr}
\end{eqnarray}
where $\tilde{\gamma} = \frac{\gamma}{p_{\rm a}}$. A simple calculation then gives the variation of the mass $m'$ of the bubble
\begin{eqnarray}
\frac{dm'}{dt} = \int_{\rm S} j_{\rm S}\,da = j_{\rm S}(r_{\rm C})\,a_{\rm S}, \label{dm'/dt}
\end{eqnarray}
where $da$ is the area measure, $a_{\rm S} = 4\pi R'^2$ the area of $\rm S$, and $j_{\rm S}(r_{\rm C})$ the value of $j_{\rm S}$ (from Eq.~(\ref{Flux-expr})) at $r = r_{\rm C}$ (and for the radius $R'$; $\rm C$ is the center of the bubble, $r_{\rm C}$ the distance $\rm OC$). Eq.~(\ref{dm'/dt}) expresses that the mass or the diameter of a bubble (the center of which is) situated at $r_{\rm C} = r_0/(1 + \frac{2\tilde{\gamma}}{R'})$ remains constant, whereas that of a bubble situated at $r_{\rm C} > r_0/(1 + \frac{2\tilde{\gamma}}{R'})$ (respectively, at $r_{\rm C} < r_0/(1 + \frac{2\tilde{\gamma}}{R'})$) decreases (respectively, increases).

\begin{figure}[b]
\begin{center}
\includegraphics[width=3.5cm]{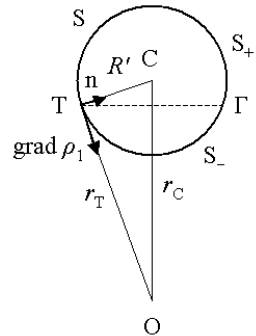}
\end{center}
\caption{Geometrical configuration. $\rm O$ is the apex of the nanoelectrode, $\rm C$ the center of the bubble, $\rm S$ its surface, $\rm T$ a point of $\rm S$ such that $\rm OT$ is tangent to $\rm S$, $\Gamma$ the horizontal circle generated by these points $\rm T$, ${\rm S}_-$ the part of ${\rm S}$ situated below $\Gamma$, and ${\rm S}_+$ that situated above $\Gamma$.} \label{Geometry}
\end{figure}

Note that, although Eq.~(\ref{S-condition}) is not strictly satisfied with the approximate expression Eq.~(\ref{rho1}), it is however ``qualitatively'' satisfied if $r_{\rm T}$ (i.e., the distance $\rm OT$) is equal to $r_0/(1 + \frac{2\tilde{\gamma}}{R'})$, $\rm T$ being a point of $\rm S$ such that $\rm OT$ is tangent to $\rm S$ (see Fig.~\ref{Geometry}).
Indeed, at $\rm T$, we have $\partial_{\rm n}\rho_1 = {\rm n}\cdot{\rm grad}\rho_1 = 0$ (since ${\rm grad}\rho_1$ is directed from $\rm T$ to $\rm O$, according to Eq.~(\ref{rho1})) and $j_{\rm S} = 0$ (since $r_{\rm T} = r_0/(1 + \frac{2\tilde{\gamma}}{R'})$), so that Eq.~(\ref{S-condition}) is satisfied at $\rm T$, i.e., on the horizontal circle $\Gamma$ of $\rm S$ generated by these points $\rm T$. Let us denote ${\rm S}_-$ the part of ${\rm S}$ situated below the circle $\Gamma$, and ${\rm S}_+$ that situated above $\Gamma$. At any point $\rm M$ of ${\rm S}_-$,  we have $\partial_{\rm n}\rho_1 = {\rm n}\cdot{\rm grad}\rho_1 < 0$ (${\rm grad}\rho_1$ being directed from $\rm M$ to $\rm O$) and $j_{\rm S} > 0$ (since $r_{\rm M} < r_0/(1 + \frac{2\tilde{\gamma}}{R'})$), so that the two members of Eq.~(\ref{S-condition}) have the same positive sign. In a similar way, the two members of this equation have the same negative sign, if $\rm M$ belongs to ${\rm S}_+$. With respect to the sign of the two members of Eq.~(\ref{S-condition}), we may then consider that this equation is ``qualitatively'' satisfied. Thus, $\rm H_2$ molecules enter the bubble through ${\rm S}_-$ ($j_{\rm S} > 0$) and leave the bubble through ${\rm S}_+$ ($j_{\rm S} < 0$). Note that the main assumption is here the approximate expression Eq.~(\ref{rho1}) (which does not take into account either the velocity field or the presence of the bubble).

\begin{figure}[t]
\begin{center}
\includegraphics[width=5.1cm]{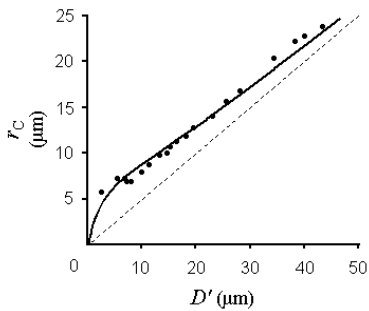}
\end{center}
\caption{Relation between $r_{\rm C}$ (distance between the center of the bubble and the apex of the electrode) and $D' = 2R'$ (diameter of the bubble) as expressed by Eq.~(\ref{rC}), with $r_0 = 9.5$ $\mu$m (continuous line), for comparison with the experimental points of Fig.~\ref{Center} (dots). The dashed line represents the equation $r_{\rm C} = D'/2$.} \label{Center-equation}
\end{figure}

A first consequence of this situation is that $r_{\rm C} > r_0/(1 + \frac{2\tilde{\gamma}}{R'})$ (since $r_{\rm C} > r_{\rm T}$) which, as noted above, indicates that the diameter of the bubble decreases with time. This is consistent with the experiments, since bubbles decreasing in diameter with time are the most frequently observed. Another consequence is the following relation between $r_{\rm C}$ and $R'$
\begin{eqnarray}
r_{\rm C} = \sqrt{R'^2 + (\frac{r_0}{1 + \frac{2\tilde{\gamma}}{R'}})^2} \label{rC}
\end{eqnarray}
(according to the value of $r_{\rm T}$; see Fig.~\ref{Geometry}), which is in agreement with the observations: see Fig.~\ref{Center-equation}, drawn with $r_0 = 9.5$ $\mu$m, for comparison with the experimental points of Fig.~\ref{Center}.
Note that, according to Eq.~(\ref{rho1}) and $\rho_1(r_0) = H\,p_{\rm a}$, this value of $r_0$ corresponds to $\dot{m_1}/M_1 = 4\pi D_1 r_0 H\,p_{\rm a}/M_1 = 4.4 \times 10^{-13}$ mol $\rm H_2$/s (produced at the apex of the nanoelectrode; $M_1$ being the molar mass of $\rm H_2$) and an effective electrolysis current intensity---averaged on the half period during which the nanoelectrode is cathode---$I_{\rm e} = 2 \times 2 N_{\rm A} q_{\rm e} \dot{m_1}/M_1 = 0.17$ $\mu$A ($N_{\rm A}$ the Avogadro constant, $q_{\rm e}$ the elementary charge). These values of $\dot{m_1}/M_1$ and $I_{\rm e}$ are divided by 2 in the general model with both $\rm H_2$ and $\rm O_2$ components.\cite{Supplement} 

From the relation $m' = \frac{M_1}{RT}\,p_{\rm a}(1 + \frac{2\tilde{\gamma}}{R'})\frac{4\pi}{3}R'^3$ between $R'$ and $m'$, and Eq.~(\ref{dm'/dt}), one easily obtains
\begin{eqnarray}
\frac{dR'}{dt} 
= K \bar H RT\,\frac{\frac{r_0}{r_{\rm C}} - (1 + \frac{2\tilde{\gamma}}{R'})}
{1 + \frac{4\tilde{\gamma}}{3R'}}, \label{dR'/dt}
\end{eqnarray}
$R$ being the gas constant, $T$ the temperature, and $\bar H = H/M_1$. If $R'$ is not too small, this expression (with $r_{\rm C}$ given by Eq.~(\ref{rC})) leads to a relatively constant value of $dR'/dt$ as $R'$ decreases. This may explain the relatively constant value of $dD'/dt$ during a large first part of each experiment (see Figs.~\ref{Diameter} and \ref{Decreasing-rate}). In addition, for large values of $R'$, i.e., also large values of $r_{\rm C}$ (according to Eq.~(\ref{rC})), Eq.~(\ref{dR'/dt}) leads to a ``universal'' limit value of $dR'/dt$
\begin{eqnarray}
(\frac{dR'}{dt})_{\lim} = -K \bar H RT, \label{dR'/dtlim}
\end{eqnarray}
which seems to be in agreement with the observations. Indeed, for large values of $D'$, we observe that $dD'/dt$ has nearly the same value, of about $-0.15$ $\mu$m/s, in very different experiments (see Fig.~\ref{Decreasing-rate}). By applying Eq.~(\ref{dR'/dtlim}), this value leads to a mass transfer coefficient $K \approx 4$ $\mu$m/s (at $20^{\circ}$C) which may be considered as a qualitatively acceptable value: indeed, in the literature, the value of this coefficient is not well known and ranges from some $\mu$m/s to some hundreds of $\mu$m/s.\cite{Itoh-Sathe:1997, Matsushima-etal:2009, Aoki-etal:2012} Note that the decreasing behavior of $dD'/dt$ for the small diameter values (Fig.~\ref{Decreasing-rate}) is not explained with this model. We currently try to improve the present model and to automate the experimental procedure in order to completely determine the conditions in which this surprising immobilization phenomenon occurs.

The presented experimental setup, i.e., the immobilized bubble and the nanoelectrode microdisplacement actuator, constitutes a point probe of acoustical pressure,\cite{Renaud-etal:2014} as well as a promising tool for sizing microbubbles.\cite{Czarnecki-etal:2015}\\ 

We thank Serge Mensah and Younes Achaoui for helpful discussions and acknowledge the financial support from ``Smart US'' ANR agreement.

\bibliography{HAL_Bubble_immobilization}
 
\end{document}